\documentclass{iauc}
\usepackage{graphicx}
\pubyear{2005}
\volume{199}
\pagerange{1--}
\jname{Probing Galaxies through Quasar Absorption Lines}
\editors{P. R. Williams, C. Shu, and B. M\'{e}nard, eds.}
\title[Bias in the H/ESO DLA survey] %% give here short title %% 
{Evidence for a magnitude-dependent bias in the Hamburg/ESO Survey for Damped Lyman--$\alpha$ Systems
}

\author[Smette et al.]   %% give here short author list %%
{A. Smette$^1$%
  \thanks{Research Associate, F.N.R.S., Belgium},
L. Wisotzki$^2$
C. Ledoux$^1$ \break 
O. Garcet$^3$
S. Lopez$^4$
\and
D. Reimers$^5$
}

\affiliation{
$^1$European Southern Observatory, Casilla 19001, Santiago, Chile,\break
email: asmette@eso.org, cledoux@eso.org\\[\affilskip]
$^2$Astrophysikalisches Institut, Potsdam, An der Sternwarte 16,
D--14882, Potsdam, Germany\break
email: lutz@aip.de\\[\affilskip]
$^3$Institut d'Astrophysique et de G\'{e}ophysique, All\'{e}e du 6 Ao\^{u}t 17,
  B--4000 Li\`{e}ge, Belgique\break
email: garcet@astro.ulg.ac.be\\[\affilskip]
$^4$Departamento de Astronomia, Universidad de Chile, 
Casilla 36-D Santiago, Chile\break
email: lopez@das.uchile.cl\\[\affilskip]
$^5$ Hamburger Sternwarte, Gojenbergsweg 112, D--21029 Hamburg, Germany\break
email: dreimers@hs.uni-hamburg.de\\[\affilskip]
}
\begin{document}

\maketitle

\begin{abstract}
We   present  preliminary results from  the   Hamburg/ESO survey for
Damped Ly-$\alpha$  (hereafter,   DLA)  Systems.   This survey    is
characterized by (i) the good knowledge  of the biases affecting the
parent QSO survey, (ii) the brightness and (iii) the relatively wide
magnitude distribution of  the   background QSOs. Therefore, it   is
well--suited to  study possible  magnitude--dependent biases  in DLA
surveys, such as the one expected from dust obscuration.

We   have  systematically  searched  for  damped   Lyman-$\alpha$ line
candidates  in 5  \AA\,
resolution  spectra of the  188  QSOs that constitute our  statistical
sample.  These candidates have later  been reobserved with UVES at the
ESO--Very Large Telescope (VLT) for  confirmation and accurate $N(HI)$
measurements. In  the redshift  range covered  by  the survey,  19 DLA
systems have been discovered. Over the  whole survey, we find that the
number  density   $n(z)$       and cosmological  density    of     gas
$\Omega_\mathrm{gas}$ have  comparable values to  the ones obtained by
CORALS (Ellison et al. 2001).

However,    the  number  densities of    DLA   systems  $n(z)$ in  two
sub--samples of equal    absorption   distance path  defined   by  the
magnitude of  the background QSOs differ  by a factor  of $\approx 5$. 
We estimate  that  the probability that   $n(z)$ is equal in  the  two
sub--samples is $< 0.003$.  A similar,  only slightly less significant
difference is found for $\Omega_\mathrm{gas}$.

%% add here a maximum of 10 keywords, to be taken form the file <Keywords.txt>.

\end{abstract}

\firstsection % if your document starts with a section,
              % remove some space above using this command.
\section{Introduction}

The possibility that dust in  intervening DLA absorbers
could obscure the background QSOs has been considered for more than 20
years (see,   e.g., Ostriker  \&  Heisler  1984).  Fall \& Pei  (1989)
estimate that  30\% of the $z_{\rm em}=3$  QSOs could have been missed
by optical  surveys  and,  conversely, that dust  could  significantly
biases surveys from  DLA systems in  the spectra of optically-selected
QSOs. Pei,  Fall \& Bechtold (1991) find  that spectra of QSOs showing
DLA absorption lines are significantly  redder than those without  DLA
systems, in an unfortunately  small sample.  Recently, Murphy \& Liske
(2004) have failed to confirm this finding in a  much larger sample of
QSOs from the SDSS survey.

Direct evidence that dust  may strongly affect  the statistics of  DLA
systems is  crucially missing  for different reasons: (i) the
QSOs from the LBQS survey (Wolfe  et al. 1995)  show a narrow magnitude
range (cf. Fig.~1); (ii)  a  comparison between  optically-  and
radio-selected DLA  samples  (Ellison et al.  2001)  does  not provide
strong  constraints  because   of    the small size  of   the
radio-selected sample. Instead,  the evidence  for dust in  DLA systems is mostly indirect (Pettini et al. 1997; Boiss\'e et al.  1998; Ledoux
et al. 2003). 
\begin{figure}
  \begin{center}
   \hbox{
     \includegraphics[height=2.8in,angle=270]{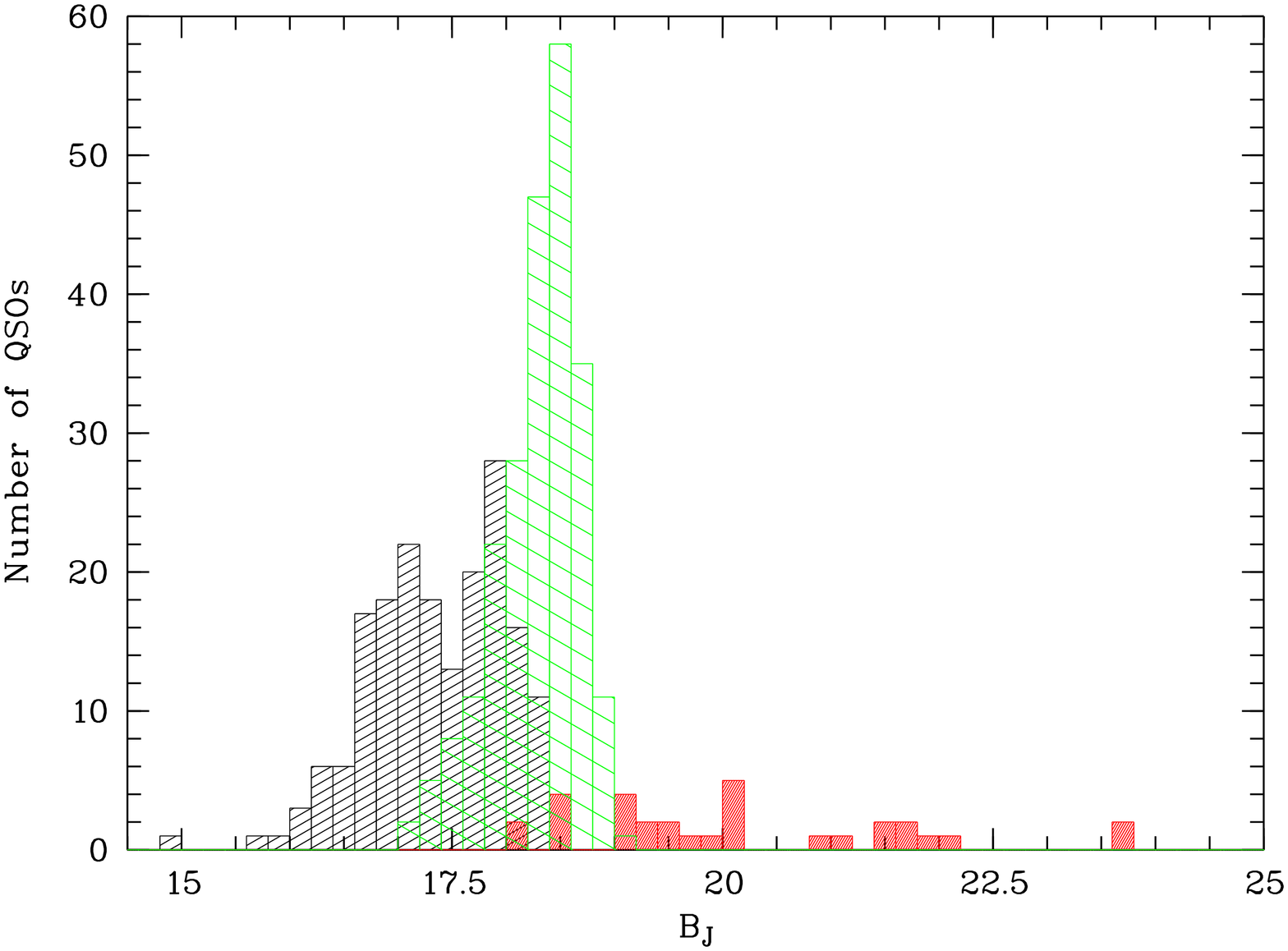}
\begin{minipage}[t]{6cm}
\vspace{3mm}
  Figure 1:  Distribution of QSO  magnitudes in  the
  three major surveys
  for DLA systems at $z_{\rm abs}\approx 2$: in \textit{Black, small
    hashing spacing: }, the recently completed Hamburg/ESO survey, in
  \textit{Red, filled histograms:} the CORALS survey (Ellison et al.
  2001), and in \textit{Green, large hashing spacing:} the LBQS
  survey, whose results are part of the study of Wolfe et al. (1995).
  Note the strong difference in $B_\mathrm{J}$ magnitude coverage
  between the three surveys.
\end{minipage}
}
\end{center}
\end{figure}
\section{A magnitude dependent bias in the Hamburg/ESO DLA Survey}
Between  1996 and 2001, we  have observed  243  QSOs or QSO candidates
from the Hamburg/ESO QSO survey at a spectral resolution of $\simeq 5$
\AA, FWHM with the ESO  -- 1.5m and Danish  -- 1.54m telescopes at ESO
La Silla observatory.  After eliminating non-QSOs, $z < 1.6$ QSOs, BAL
QSOs, or spectra    with a too  low   S/N, we built  a   statistically
well-defined sample of 188 QSOs. The distribution of these QSOs in the
$z - B_\mathrm{J}$ plane  is statistically undistinguishable from  the
parent H/ESO QSO survey,   indicating that we have  not  unvoluntarily
biased our surveys while selecting our spectroscopic targets.

The spectra of   these  objects   were systematically   searched   for
absorption  lines in  a redshift range   limited, in the blue,  by the
wavelength for which the  S/N per pixel is  5 and, in  the red, by the
wavelength corresponding to a receding velocity  of 5000 km/s from the
QSO redshift.   The QSOs whose spectra  show absorption lines detected
at the 5$\sigma$ level with rest equivalent width larger than 7.5\AA\,
(corresponding to  $\log{N_\mathrm{HI}} =  20$)  were reobserved  with
UVES at the ESO VLT.  Results for  the whole survey are given in
the 2nd column   of  Table   1    ($\Omega_\mathrm{M}   = 1.0$,
$\Omega_\Lambda = 0$ for easy comparison with previous results).
In other  words,  our mean  number  density for   the whole survey  is
extremely similar to  the one of CORALS  (Ellison et al. 2001), if we
take the redshift dependence  of  $n(z)$ into account,   and is 50  \%
larger  than  that   found   from the   relation   $n(z) =  0.055   \,
(1+z)^{1.11}$ (Storrie--Lombardi \& Wolfe 2000).    We also find   the
cosmological density of neutral gas comparable to previous results.

\begin{figure}
  \begin{center}
    \hbox{
    \includegraphics[height=2.3in,angle=90]{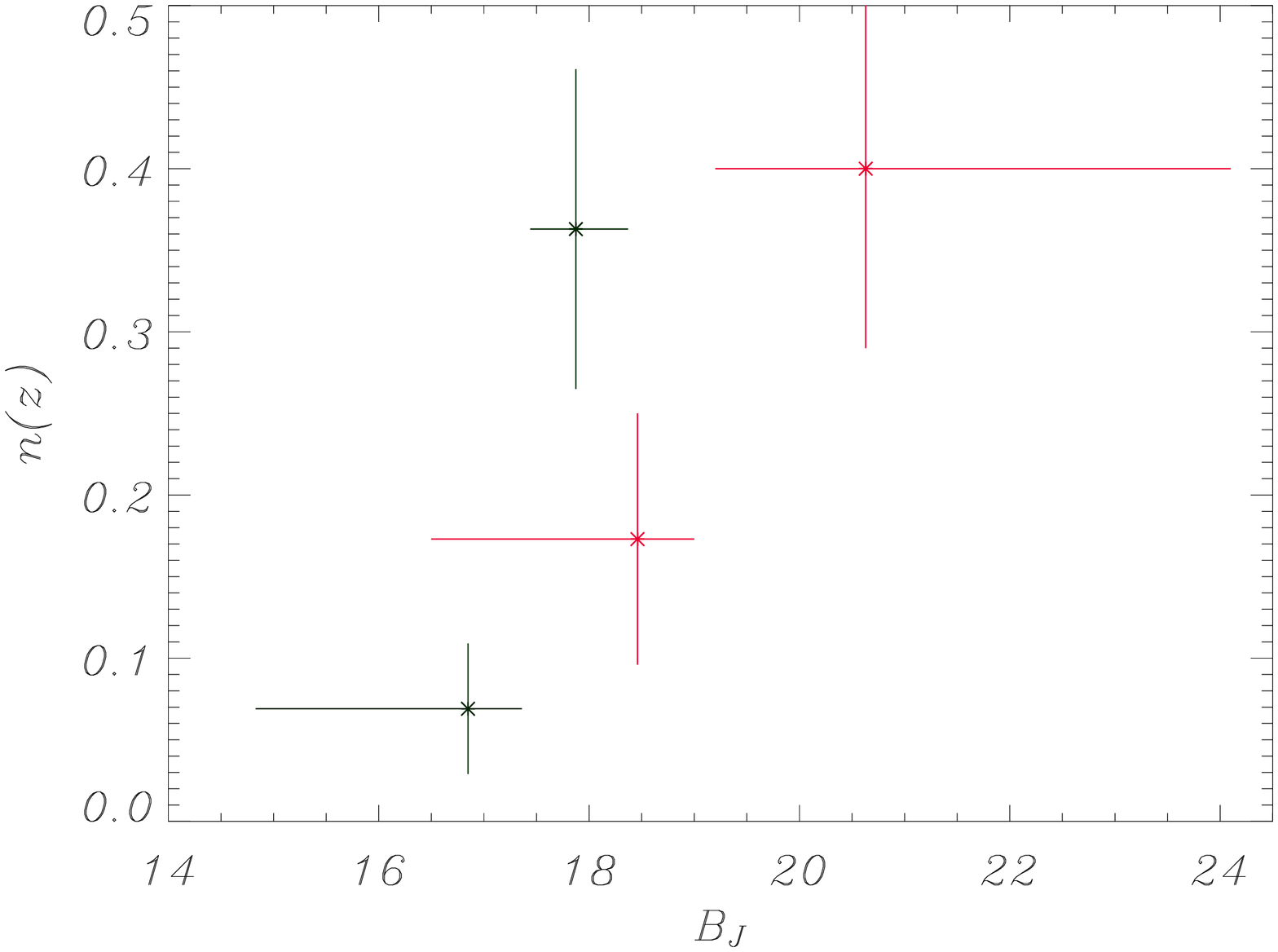}
    \includegraphics[height=2.3in,angle=90]{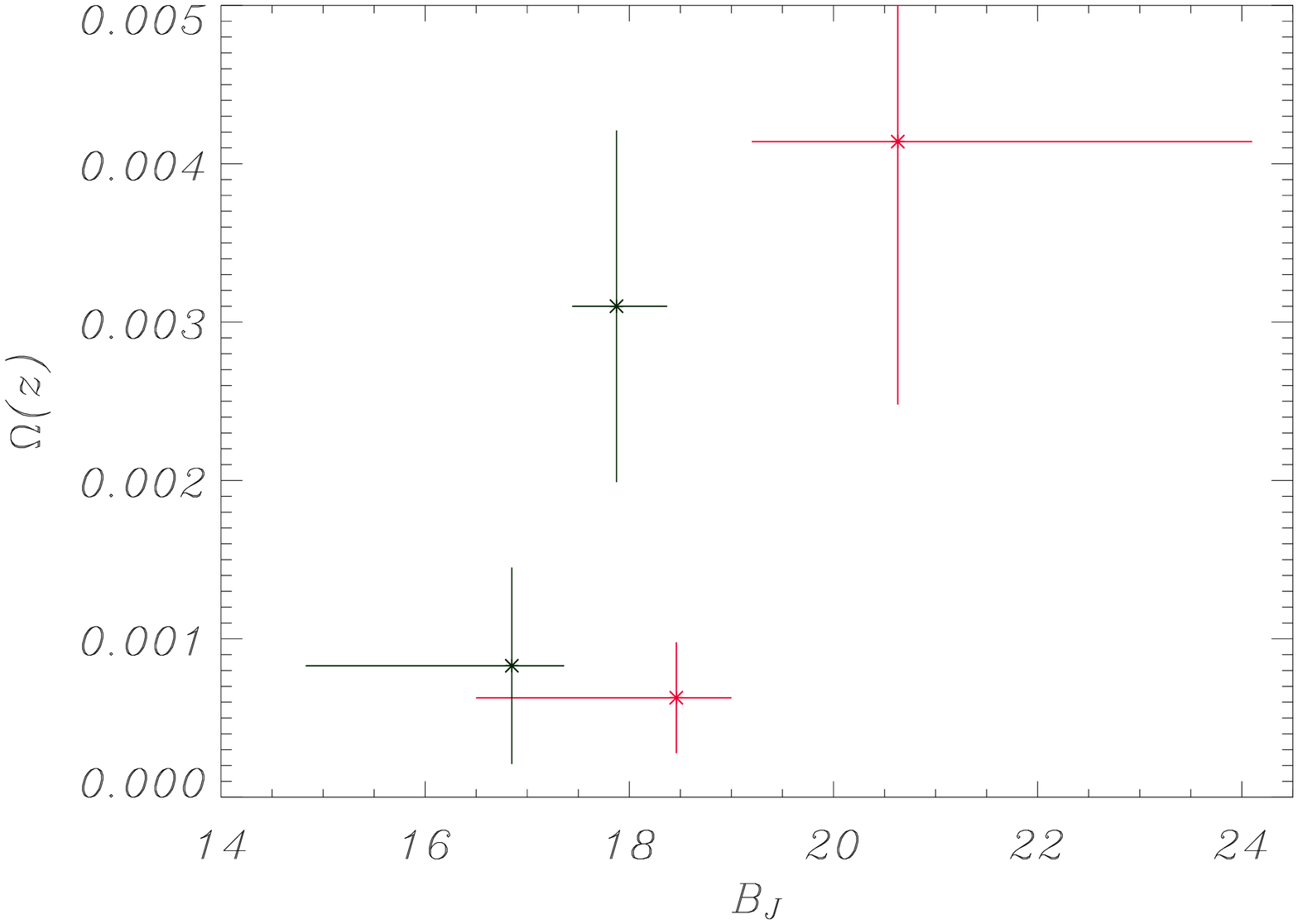}
  }
\caption{\textit{Left:} Number densities of  DLA systems in
  the bright and faint sub--samples of the H/ESO survey 
(black, left--most points) and CORALS (red, right--most
points). \textit{Right: }
  Idem, for the cosmological density. 
The 2 CORALS sub--samples have  identical $\Delta  \chi$: the bright
  sample  correspond to a coverage   towards $B_\mathrm{J} < 19.2$
  QSOs and
  include 5 DLA systems, while the faint sample corresponds to fainter QSOs
  and      include 12 systems.   The   horizontal   bars represent  the
  $B_\mathrm{J}$  coverage  of each sub--sample.    The position of the
  vertical  bar  along the   $B_\mathrm{J}$  axis represents the  mean
  magnitude  of  the sub--sample.   The   extent  of the vertical  bar
  represents the $\pm1\sigma$ error.}
\end{center}
\end{figure}

%\section{A magnitude--dependent bias}

Our QSO sample has  the particular characteristics of including bright
QSOs.  We  have therefore  constituted  2 sub--samples of  equal total
absorption distance path  $\Delta \chi$.  These sub--samples are based
on the magnitude $B_\mathrm{J}$  of the background QSOs.  The critical
magnitude  to define  the 2 samples  is $B_\mathrm{J}  =  17.40$. From
Poisson statistics, we  estimate that  the  probability to  obtain the
observed number of DLA  systems in  the  2 sub--samples from the  same
(mean) number density is   less than 0.003.  Figures   2    show a
comparison of the values of $n(z)$ and $\Omega_\mathrm{gas}$ for the 2
H/ESO sub--samples with 2 sub--samples of the CORALS survey defined in
a similar way.
\begin{table}[ht]\def~{\hphantom{0}}
  \begin{center}
    \caption{Preliminary results from the H/ESO DLA survey}
              \label{tab:results}
    \begin{tabular}{cccc}\hline      
                & Whole  & Bright & Faint  \\
                & Sample & Sub--Sample & Sub--Sample \\
                \hline
                QSOs         & 188            &  93    & 95    \\
                $\Delta z$      & 87.7           &  43.6  & 44.0  \\
                $\Delta \chi$   & 154.1          &  76.4  & 77.8  \\
                $<B_\mathrm{J}>$&  17.37         &  16.85 & 17.87 \\
                \hline
                DLAs         & 19             &  3     & 16 \\
                $n(z) $         & $0.22\pm0.05$ & $0.069\pm0.040$  & $ 0.36\pm0.09$\\
                $10^{3} \, \Omega_\mathrm{gas}$ & $1.98\pm0.64$ & $0.83\pm0.62$&$3.10\pm1.12$\\
              \hline
              \end{tabular}
            \end{center}
\end{table}

Both the H/ESO and CORALS surveys indicate that $n(z)$ and
$\Omega_\mathrm{gas}$ are correlated with the magnitude of the background
QSOs. 
The steep increase seen in both surveys does not take place at the
same $B_\mathrm{J}$ though. A possible reason is that the CORALS QSOs
have larger redshifts than the H/ESO ones. Indeed, 
(a) for (nearly) all H/ESO QSOs, the Ly--$\alpha$ emission line falls
within the $B_\mathrm{J}$ passband; this is not the case for  a
significant fraction of the CORALS QSOs; (b) the flux decrement caused
by the Ly-$\alpha$ forest is larger for the CORALS QSOs than for the
H/ESO ones; (c) as the mean redshift of the CORALS survey is larger, any
reddening effect caused by the DLA systems would be larger in the
CORALS survey
than in the H/ESO one. We are currently testing this possibility.

%\begin{acknowledgments}
%\end{acknowledgments}

\end{document}